\documentclass[preprint2]{aastex}
\usepackage{color}

\shorttitle{Pulsations before the Onset of Solar Flares}
\shortauthors{Tan}

\begin{document}

\title{Very Long-period Pulsations before the Onset of Solar Flares}

\author{Baolin Tan$^{1,3}$, Zhiqiang Yu$^{2,3}$, Jing Huang$^{1,3}$, Chengming Tan$^{1,3}$, and Yin Zhang$^1$}

\affil{$^1$Key Laboratory of Solar Activity, National Astronomical
Observatories of Chinese Academy of Sciences, Beijing 100012,
China; bltan@nao.cas.cn}

\affil{$^2$ Applied Physics Department, Harbin Institute of
Technology, Harbin 15001, Heilongjiang, China}

\affil{$^3$ School of Astronomy and Space Sciences, University of
Chinese Academy of Sciences, Beijing 100049, China}

\begin{abstract}

Solar flares are the most powerful explosions occurring in the
solar system, which may lead to disastrous space weather events
and impact various aspects of our Earth. So far, it is still a big
challenge in modern astrophysics to understand the origin of solar
flares and predict their onset. Based on the analysis of soft
X-ray emission observed by the \emph{Geostationary Operational
Environmental Satellite} (GOES), this work reported a new
discovery of very long-periodic pulsations occurred in the
preflare phase before the onset of solar flares (preflare-VLPs).
These pulsations are typically with period of 8 - 30 min and last
for about 1 - 2 hours. They are possibly generated from LRC
oscillations of plasma loops where electric current dominates the
physical process during magnetic energy accumulation in the source
region. The preflare-VLP provides an essential information for
understanding the triggering mechanism and origin of solar flares,
and may help us to response to solar explosions and the
corresponding disastrous space weather events as a convenient
precursory indicator.

\end{abstract}

Online-only material: color figures

\section{Introduction}

Solar flare is a sudden, rapid, and violent magnetic-energy
release and brightness enhancement in broad spectrum of emissions
observed in the solar atmosphere near sunspots (Shibata and Magara
2011). Generally, solar flares are classified into A-, B-, C-, M-,
and X-class according to their maximum flux at soft X-ray (SXR)
wavelength of 1.0 - 8.0 \AA~ measured by the standard
\emph{Geostationary Operational Environmental Satellite} (GOES, a
satellite series initially deployed in 1974, GOES-14 and GOES-15
are inline during the present solar cycle 24) (Tandberg-Hanssen \&
Emslie 1988). A typical X-class flare may release magnetic-energy
of more than $10^{25}$ J into the interplanetary space and impact
greatly upon various aspects of our Earth (Nonweiler 1958,
Scafetta and West 2003, Pick and Vilmer 2008, Gonzalez et al.
2014). Although there is a general agreement on the flares' causes
of magnetic-reconnection in solar atmosphere, there are still many
big unclear problems: What is the trigger of magnetic-reconnection
in the source region? How does the enormous magnetic-energy
transform into kinetic energy carried by particles and plasmas? In
particular, how do we predict a powerful solar flare and the
related disastrous space weather event? Naturally, it is very
important and convenient to find out the easily-obtained and
confirmatory precursors of solar flares. So far, many clues are
reported from multi-wavelength observations in preflare phases,
including radio spectral fine structures (Zhang et al. 2015),
filament activities (Chifor et al. 2006), weak SXR bursts (Tappin
1991), magnetic helicity accumulations (Zhang, Tan, \& Yan, 2008),
etc. However, some of them are very weak with remarkable
uncertainties, and some of them are very complicated and difficult
to recognize from the huge ocean of observation data (Martin 1980,
Bloomfield et al. 2012).

Generally, a solar flare can be partitioned into three phases
according to its GOES SXR flux light curve at wavelength of 1.0 -
8.0 \AA~: preflare (before the flare onset), rising (from the
flare onset to its maximum, the time interval is named rising-time
of the flare) and postflare (after the flare maximum). A typical
flare event and its phase partition is shown in Figure 1.

\begin{figure}[ht] % Figure 1
\begin{center}
   \includegraphics[width=8.0 cm]{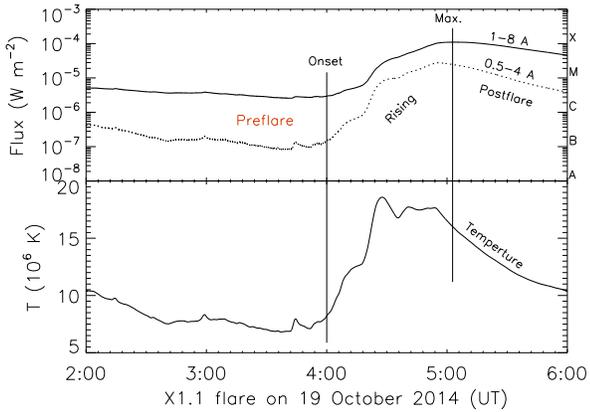}
\caption{Flare classification and the phase partition. The upper
panel is the soft X-ray light curves at 1 - 8 \AA~ and 0.5 - 4
\AA, while the bottom presents the profile of the corresponding
temperature.}
\end{center}
\end{figure}

Because the GOES program consists of a series of geostationary
satellites which overlap in time so that there are always one to
three spacecrafts working in orbit, and this guarantees an
essentially uniterrupted time series of recorded solar SXR fluxes
(Thomas, Starr, and Crannell 1985, Garcia 1994, White, Thomas, and
Schwarts 2005). In this work, we mainly investigated the GOES SXR
observation data during preflare phase, and found that there were
very long-period pulsations (VLPs) with typical period of 8 - 30
min and typical duration of about 1 - 2 hours occurring just
before the onset of solar flares. We named this phenomenon as
preflare-VLP. As we know that VLPs with similar timescales are
frequently reported during flare processes or even non-flaring sun
(Harrison 1987, Svestka 1994, Wang 2011, Tan et al. 2010, Yuan et
al. 2011, etc.). However, the preflare-VLPs reported in this work
are the first time observed just in the preflare phase of solar
flares, and may be meaningful to reply the questions mentioned in
the above paragraphs.

This paper is organized as following. Section 2 introduces the
main properties of preflare-VLPs in several typical flare events
and section 3 presents the statistic characteristics of solar
flares in the solar cycle 24. The physical mechanism and the
related theoretical discussions are presented in section 4, and
finally, some conclusions are summarized in section 5.

\section{Several Typical Flare Events Accompanying with Preflare-VLPs}

In order to eliminate the influence of other flares, this work
focused on analyzing the isolated flares in the solar cycle 24
observed by GOES at SXR wavelengths of 1.0 - 8.0 \AA~ and 0.5 -
4.0 \AA~ with cadence of 2 seconds, which in principle probe the
dynamical process of energy release in flaring active regions
(Tandberg-Hanssen \& Emslie 1988, Georgoulis, Vilmer, and Crosby
2001). Here, an isolated solar flare is defined as that there is
no same or higher class of flare event occurring in 2 hours before
the flare onset, and no saturation or bad recorded data to disturb
the analyzing results.

Figure 2 presents SXR light curves at 1.0 - 8.0 \AA~ and 0.5 - 4.0
\AA~, and the related temperatures in 4 typical flare events. The
temperature is derived from the ratio of SXR intensities at the
above two wavelength bands (Thomas, Starr, and Crannell 1985,
White, Thomas, and Schwarts 2005). All background emission is
subtracted from SXR flux intensity. By a simple view, we find that
there are trains of pulses with approximately equal time-intervals
occurred before the onset of the corresponding flare event. We
call such train of pulses as preflare very long-period pulsation
(\textbf{preflare-VLP}). Here, it is necessary to define a
detected criterion of preflare-VLP: (1) occurred during 2 hours
before the flare onset, (2) lasted for more than 30 minutes
(duration, $D>$ 30 min) and composed of at least 4 pulses, (3) the
maximum amplitude of each pulse is higher than 2$\sigma$ ($\sigma$
is the standard derivation of the background temperature before
the train of pulses), and (4) the time-interval between adjacent
pulses is called period ($P$), and the maximum period is shorter
than 2 times of minimum period ($P_{max}<2P_{min}$) and $P>$ 1
min. When there is a pulsation satisfying the above criterions in
2 hours before the flare onset, we say the flare accompanying with
preflare-VLP, and when there is no such pulsation we say the flare
is without preflare-VLP.

\begin{figure*}[ht] % Figure 2
\begin{center}
   \includegraphics[width=15 cm]{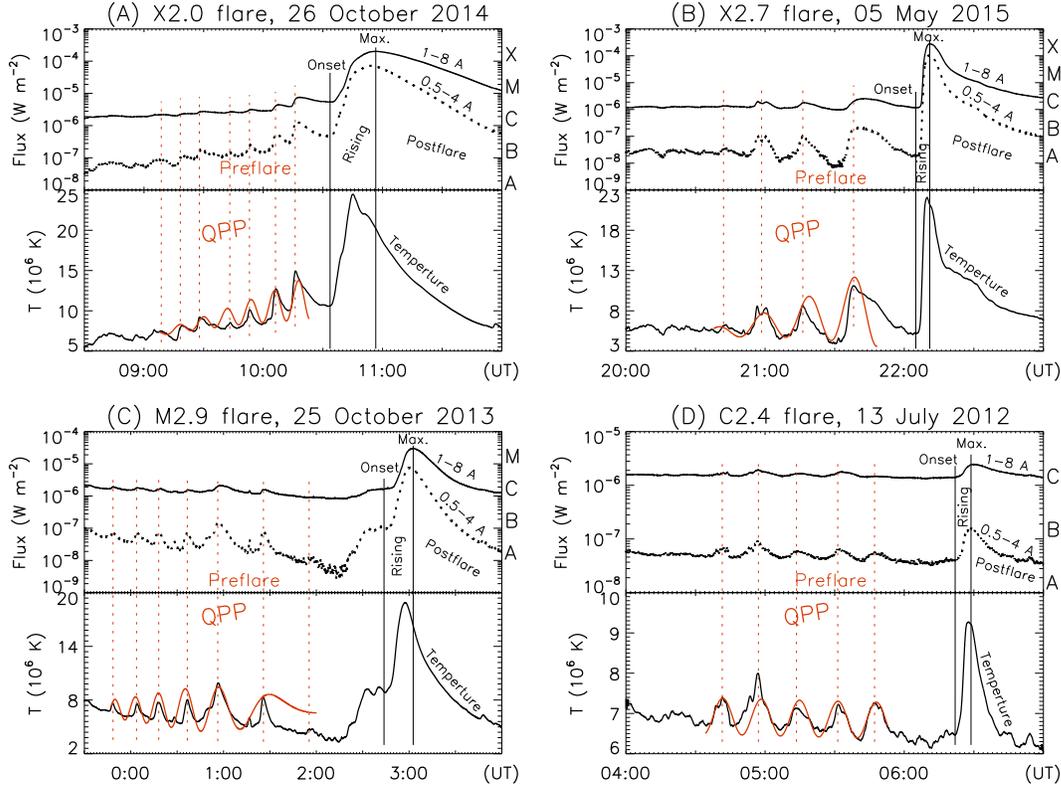}
\caption{Very long-period pulsations before the onset of solar
flares (preflare-VLP) in four typical solar flares. In each event,
the upper panel shows light curves of soft X-ray emission flux at
wavelength of 1.0 - 8.0 \AA~ and 0.5 - 4.0 \AA~, while the bottom
panel shows profile of the corresponding temperature. The
observations are obtained by GOES-15. The vertical dotted lines
marked pulses of the preflare-VLP.}
\end{center}
\end{figure*}

\subsection{X2.0 flare on 26 October 2014}

Figure 2A shows the process of an X2.0 flare occurred on 26
October 2014. The flare took place in a super active region
NOAA12192 located S14W37 closed to the center of solar disk. The
flare started at 10:35 UT, reached to maximum at 10:56 UT, and the
rising-time was about 21 min. During the two hours preflare phase
before the flare onset, from 08:30 UT to 10:35 UT, there was no
flare stronger than M-class. Therefore the X2.0 flare is an
isolated flare event.

During the preflare phase of the X2.0 flare, there is a
preflare-VLP which contains 7 pulses and lasts for about 80
minutes. The pulse period is in the range of 9.5 - 13.0 min, and
the average period is about 11.4 min (684 s). It is very
interesting that the pulse amplitudes are increased slowly from
about 1.0 MK to 4.0 MK at the temperature profile. The
preflare-VLP can be fitted by a cosine function from a
least-square method,
\begin{equation}
T\approx5.0+10^{-3}t+6.1\times10^{-8}t^{2}\cos(\frac{2\pi t}{684})
\end{equation}
Here, T is the temperature in unit of $10^{6}$ K (MK), t is time
in unit of second and start from 08:30 UT. The constant 5.0 MK is
the average temperature of the background before the preflare-VLP.
The second term $10^{-3}t$ represents the slow changes of the
background during the preflare-VLP. The function
$6.1\times10^{-8}t^{2}$ in the third term shows the changes of the
amplitude of the pulsation. Here, the period of the pulsation is
near a constant (684 s).

\subsection{X2.7 flare on 05 May 2015}

Figure 2B shows an X2.7 flare occurred on 05 May 2015 in active
region NOAA12339 located N15E79, very close to the east limb of
solar disk. The flare started at 22:05 UT, rapidly reached to its
maximum at 22:10 UT, and the rising-time was only 5 min. During
the preflare phase from 20:00 UT to 22:05 UT, there was also no
flare stronger than M-class. Therefore this flare is an isolated
flare event.

During the preflare phase, there is a preflare-VLP which contains
4 pulses and lasts for about 70 min. The pulse period is in range
of 16.0 - 20.0 min, and the average period is about 18.4 min (1104
s). Similarly, the pulse amplitudes are increased from about 0.5
MK to 5.5 MK at the temperature profile. We can also apply a
cosine function from a least-square method to fit the
preflare-VLP,
\begin{equation}
T\approx3.5+8.0\times10^{-4}t+1.2\times10^{-7}t^{2}\cos(\frac{2\pi
t}{1104}).
\end{equation}
Here, t starts from 20:00 UT. This fitted function is very similar
to Equation (1), its average temperature of the background before
the preflare-VLP is only about 3.5 MK.

\subsection{M2.9 flare on 25 October 2013}

Figure 2C shows an M2.9 flare occurred on 25 October 2013 in
active region NOAA11882 at S07E76 closed to the east limb of solar
disk. The flare started at 02:48 UT, peak at 03:02 UT, and the
rising-time is about 14 minutes. It is also an isolated flare.

Before the flare onset, there is a preflare-VLP containing 7
pulses and lasting for about 130 min. The pulse period is
increased from 15 min up to 30 min, and the average period is
about 21.2 min. The pulse amplitude increased in the first half
and then decreased slowly in the second half of the preflare-VLP.
The preflare-VLP can be also fitted by a complex cosine function,
\begin{equation}
T\approx7.0-5\times10^{-5}t+(\frac{5000}{t-4800})^{2}\cos[\frac{(18200-t)\pi}{2.88\times10^{6}}(t+200)].
\end{equation}
t starts from 23:30 UT on 24 October 2013. Here, the average
temperature of the background before the preflare-VLP is 7.0 MK,
and the change of the background temperature during the
preflare-VLP is negative, opposite to Equation (1) and (2). At the
same time, the period and amplitude of the pulsation have more
complicated changes with time.

\subsection{C2.4 flare on 13 July 2012}

Figure 2D shows a small C2.4 flare occurred on 13 July 2012 in
active region NOAA11515 at S19E10 near the center of solar disk.
The flare started at 06:22 UT, maximum at 06:29 UT. The
rising-time is about 7 min. It is still an isolated flare event.

During the preflare phase, a preflare-VLP is occurred with 5
pulses and lasted for about 75 min. The pulse period is in the
range of 15.0 - 17.0 min with average of 16.4 min (984 s). We can
also use a cosine function to fit the preflare-VLP,
\begin{equation}
T\approx7.0-2.5\times10^{-5}t+0.45\cos(\frac{2\pi t}{984}+1.5).
\end{equation}
t starts from 04:00 UT on 13 July 2012. Similar to Equation (3),
here, the average temperature of the background before the
preflare-VLP is about 7.0 MK, and the change of the background
temperature during the preflare-VLP is negative. Both the
amplitude and period are near constant.

Furthermore, Figure 2 also shows that the preflare-VLP at 1.0 -
8.0 \AA~ is fully in-phase with that at wavelength of 0.5 - 4.0
\AA~ as well as at the corresponding temperature profiles. The
amplitude of temperature variations is in the range of 0.5 - 5.5
MK. This fact indicates that the temperature of flaring region
also experiences quasi-periodic variations before the flare onset.

In the above descriptions, we just presented the results derived
directly from counting the pulses of preflare-VLP. It is displayed
very simple and straightforward. Actually, in order to check the
results, we also applied two mathematic analytic methods of fast
Fourier transformation (FFT) and wavelet transformation (WLT) to
the time-series data of temperature, and found the results are
agree with each other.

\section{Statistic Analysis of the Preflare-VLP in Solar Cycle 24}

In order to demonstrate the universality of the preflare-VLP, we
made a statistic investigation of isolated solar flares. By using
the detected criterion described in Section 2, we distinguished
totally 412 isolated flares in the Solar Cycle 24 (from October
2010 till to July 2016), including 39 X-class, 183 M-class, and
190 C-class flares. We applied three different methods to
distinguish the emission periodicity in the preflare phase of the
above flare events: (1) directly counting the pulses, (2) fast
Fourier transformation, and (3) wavelet transformation, and
authenticate each other in a secure manner.

Among the 412 isolated flares, there are 144 flares ($\sim$ 35\%)
having preflare-VLPs, including 18 X-class (46\%), 76 M-class
(42\%), and 50 C-class (26\%) flares. The other 268 flares have no
obvious preflare-VLP. Table 1 presents the statistic results of
the period and duration of preflare-VLP, and the rising-time of
the isolated flares. Figure 3 presents their distributions with
respect to the flare class.

\begin{deluxetable}{cccccccccccc} % Table 1
\tablecolumns{10} \tabletypesize{\scriptsize} \tablewidth{0pc}
\tablecaption{Preflare-VLP statistics of the isolated flares in
solar cycle 24\label{tbl-1}} \tablehead{
 \colhead{Type}      & \colhead{parameter} & \colhead{X-class}           & \colhead{M-class}          & \colhead{C-class}         \\}
\startdata
with preflare-VLP    &  Number             &   18(46\%)                  &     76(42\%)               &    50(26\%)               \\
                     &  Period (min)       &  4.4 - 47.3 (16.1$\pm$10.7) &  1.9 - 46.2 (16.2$\pm$7.9) & 7.7 - 33.3 (16.4$\pm$7.4) \\
                     &  Duration (min)     &  45 - 180 (85.6$\pm$33.7)   &  17 - 200 (98.5$\pm$35.8)  & 42 - 185 (97.4$\pm$33.2)  \\
                     &  Rising-time (min)  &  4 - 33 (14.7$\pm$9.1)      &  2 - 47 (11.7$\pm$9.5)     & 3 - 36 (9.8$\pm$8.3)      \\\hline
without preflare-VLP &  Number             &   21(54\%)                  &     107(58\%)              &    140 (74\%)             \\
                     &  Rising-time (min)  &  6 - 95 (25.2$\pm$23.0)     &  2 - 103 (16.7$\pm$16.5)   & 3 - 78 (13.2$\pm$11.8)    \\
\enddata
\tablecomments{The format of the numerical value of period,
duration, and rising-time is: minimum - maximum (average $\pm$
mean square deviation). The percentage in the parentheses is the
proportion in the same flare class.}
\end{deluxetable}

\begin{figure*}[ht] % Figure 3
\begin{center}
   \includegraphics[width=12 cm]{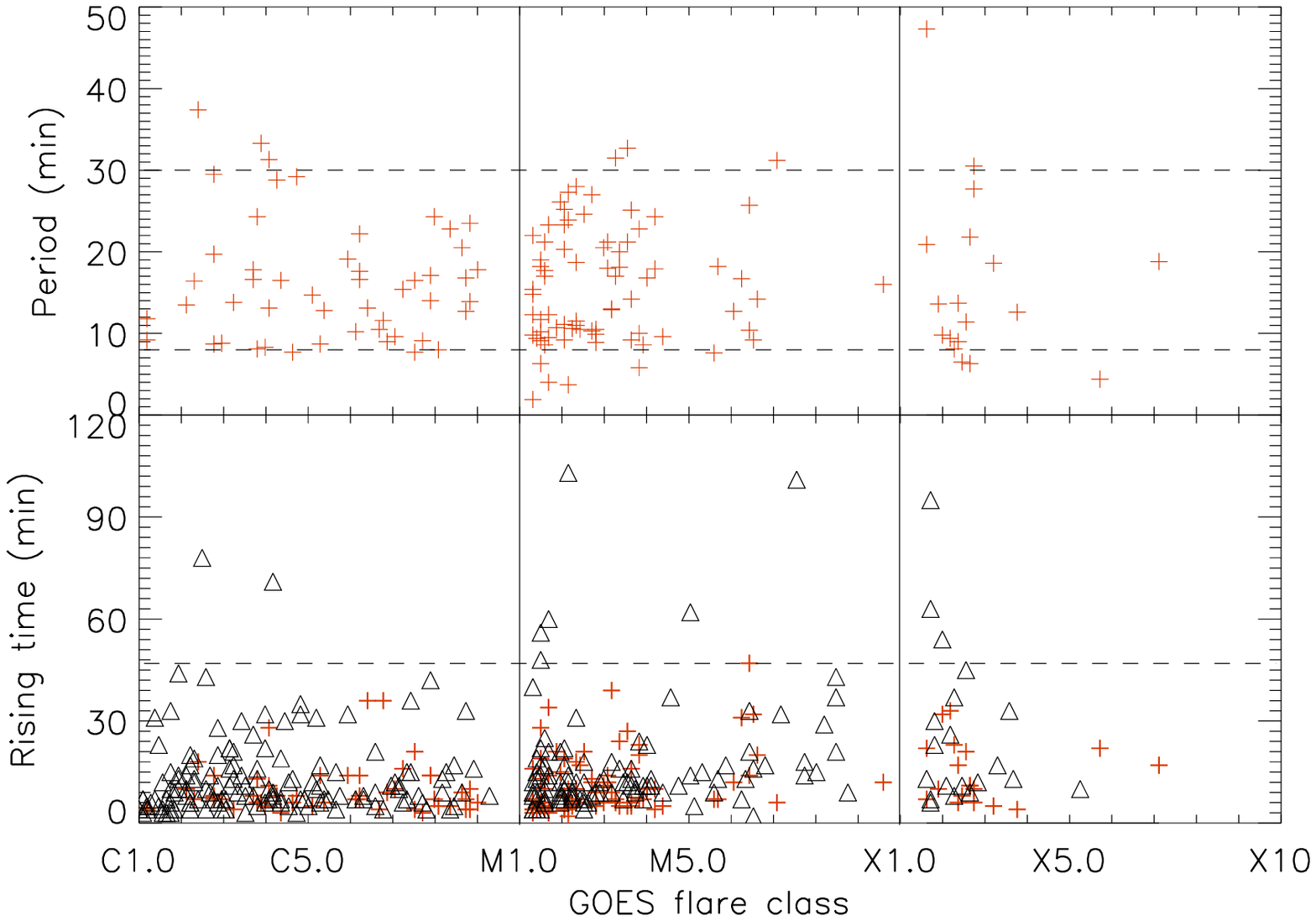}
\caption{Statistics of the isolated C- (left), M- (middle) and
X-class flares (right) in solar cycle 24. The upper panels are
period distribution of the preflare-VLPs, the bottom panels show
the rising-time distribution of the isolated flares. Red pluses
(+) indicate flares with preflare-VLPs while triangles
($\triangle$) indicate flares without preflare-VLP.}
\end{center}
\end{figure*}

From Table 1 and Figure 3, we can obtain several interesting
conclusions:

(1) Each preflare-VLP lasts for about 30 - 200 min (the typical
duration is in the range of 60 - 120 min, i.e. 1 - 2 hours) and
contains 4 - 11 pulses. The typical pulsating period is in the
range of 8 - 30 min. The longest period is 47 min (X1.0 flare, 11
June 2014) and the shortest one is 1.9 min (M1.0 flare, 5 November
2013). The averages are 16.4$\pm$7.4, 16.2$\pm$7.9, and
16.1$\pm$10.7 min in C-, M-, and X-class flares, respectively,
just around a typical value of 16$\pm$8 min. Here, the values
following the symbol $\pm$ are statistic mean square deviations.

(2) There is almost no obvious correlation between the timescales
($P$ and $D$) of preflare-VLPs and the GOES SXR classes of the
host flares. The difference of the average periods is very small
between different flare classes.

(3) The stronger flares have higher probability to produce
preflare-VLPs as well as the flares with shorter rising-time. For
example, there are about 46\% X-class flares have preflare-VLPs
while the same percentage is only 26\% in C-class flares. The
average rising-times of C-, M-, and X-class flares with
preflare-VLPs are 9.8$\pm$8.3, 11.7$\pm$9.5, and 14.7$\pm$9.1 min,
while the same values for without preflare-VLPs are 13.2$\pm$11.8,
16.7$\pm$16.5, and 25.2$\pm$23.0 min, respectively. The longest
rising-time of flare with preflare-VLP is shorter than 48 minutes,
while all the flares with rising-time above 50 min are lack of
preflare-VLP.

In fact, Figure 1 presents a typical example of long-rising
X-class flare without preflare-VLP. The rising-time is 63 min.
There is no obvious pulse during 2 hours in the preflare phase.
Long-rising flares may release energy relatively slow in complex
multi-loop system, and the plasma loops may disturb to each other,
which drive the dynamic processes to become very complicated,
obscure the signal of preflare-VLPs from the observations.

In section 2, we applied some cosine functions from the
least-square method to fitting the preflare-VLP in Equation (1) -
(4). In fact, we may summarize the fitted function into an unified
format:
\begin{equation}
T=T_{0}+bt+M(t)\cos(\frac{2\pi}{P}t).
\end{equation}
Here, $T_{0}$ is the average temperature of the background active
region before the preflare-VLP.

$b$ is the growth rate of the average temperature of the
background during preflare-VLP. It is positive ($b>0$) in most
flare events, such as the X2.0 flare on 26 October 2014 and X2.7
flare on 05 May 2015. However, it is amazing that $b$ is negative
in some events, such as the M2.9 flare on 25 October 2013 and C2.4
flare on 13 July 2013 ($b<0$). Statistics indicates that there are
about 20\% flare events with decreasing mean temperature during
the preflare phase $b<0$.

$M(t)$ is the pulse amplitude of the preflare-VLP which is
generally a function of time. In the X2.0 flare on 26 October
2014, $M(t)=6.1\times10^{-8}t^{2}$ which is increasing with time.
In the M2.9 flare on 25 May 2013 $M(t)=(\frac{5000}{t-4800})^{2}$.
But in the case of C2.4 flare on 13 July 2012, $M(t)=0.45$ is
approximated to a constant.

$P$ is the period of preflare-VLP which is generally a constant in
most events, for example, $P=684$ s in the X2.0 flare on 26
October 2014, $P=1104$ s in the X2.7 flare on 05 May 2015, and
$P=984$ s in the C2.4 flare on 13 July 2012. However, in the case
of M2.9 flare on 25 October 2013,
$P=\frac{5.78\times10^{6}}{18200-t}$ is increasing with time.

\section{Generation Mechanism of Preflare-VLP}

Then, what is the cause of the formation of VLPs during the
preflare phase?

As we know, it is ubiquitous that quasi-periodic pulsations (QPPs)
with periods from sub-second to several minutes in
multi-wavelength observations occur during the flare rising and
postflare phases (Aschwanden 1987, Tan et al. 2010, Kupriyanova et
al. 2010, Simoes, Hudson, and Fletcher 2015, etc.). The
magnetohydrodynamic (MHD) oscillations are generally applied to
explain the formation of QPPs. Because MHD oscillations can affect
almost all aspects of the emission processes: magnetic
reconnection and modulation of its rate, electron acceleration and
dynamics, and plasma conditions. Periods and other parameters are
linked with properties of emitting plasmas and morphology of
magnetic fields (Roberts, Edwin, and Benz 1984, Aschwanden 1987,
Nakariakov and Melnikov 2009). Foullon et al. (2005) reported
long-period pulsations with timescale of 8 - 12 min of X-ray
radiation during solar flares and interpreted them as a periodic
pumping of electrons in a compact flaring loop modulated by MHD
oscillation. Some people suggested that such pulsations might be
related to the slow-mode oscillations in large scale coronal loops
(Svestka 1994), while other work suggested that the long-period
pulsations could be associated with gravity-driven solar interior
modes and connected with the wave leakage of chromospheric
oscillations (Yuan et al. 2011). Another possible explanation of
VLPs is the thermal overstability of standing slow magnetoacoustic
waves in the magnetic flux loops (Kumar, Nakariakov, and Moon
2016), such as the SUMER oscillations with periods in the range of
5 - 40 min and detected in long loops with lengths of 200 - 300 Mm
(Wang 2011).

The MHD oscillation mode should be the most favorable candidate
for explaining the formation of VLPs. However, when we utilize
this mechanism to demonstrate the formation of preflare-VLPs we
meet some serious questions: the flare explosion may be the
trigger of VLPs during the flare, but what is the trigger of the
oscillations before the flare onset without explosions? and how do
we make a natural connection with the details of energy
accumulation in preflare phase?

Actually, the preflare-VLP is much alike the quasi-periodic
oscillations in precursor phase of a Tokamak¡¯s major disruption
(Wesson 1997, Jiang et al. 2015). The precursor phase (similar to
the preflare phase) lasts for typically 10 ms and the fast phase
(similar to the flare rising phase) for about 1 ms at medium-sized
Tokamaks. Because of electric currents and limited resistivity,
the tearing-mode instability can produce the growth of magnetic
oscillations of an m=2 mode in the precursor phase of Tokamak
major disruptions (Jiang et al. 2015).

Similar to Tokamaks, the solar flaring loops also have
longitudinal electric currents (Tan and Huang 2006, Tan et al.
2006). During the preflare phase, the flaring source region
accumulates magnetic energy gradually through the photospheric
convection. The photospheric convection can drive shearing,
rotating and twisting motions around loop footpoints and drive
electric currents in the plasma loop (Alfven and Carlqvist, 1967,
Tobias and Cattaneo, 2013). The current-carrying plasma loop is
analogous to a LRC-circuit with electric inductance
$L=\frac{\mu_{0}l}{\pi}(\ln\frac{8l}{\sqrt{\pi S}}-\frac{7}{4})$
and capacitance $C=\frac{8\pi\rho S^{2}}{\mu_{0}^{2}lI^{2}}$.
Here, $\rho=n_{e}m_{e}+n_{i}m_{i}\approx nm_{i}$ is the plasma
density in unit of $kg\cdot m^{-3}$, $n$ is the plasma number
density, $S$, $l$ and $I$ are the cross-sectional area ($m^{2}$),
length of the plasma loop ($m$) and and electric current ($A$),
respectively. Such LRC-circuit will produce intrinsic oscillation
(Zaitsev et al. 1998) with period,

\begin{equation}
P=2\pi\sqrt{LC}\approx2.75\times10^{4}\frac{S\sqrt{\rho}}{I}.
\end{equation}

LRC-oscillation can modulate both thermal and nonthermal emission.
The thermal emission contributes to SXR emission while the
nonthermal emission contributes to hard X-ray emission and
energetic particles. The period of LRC oscillation is proportional
to the cross-sectional area (S), and anti-proportional to the
electric current (I) in the loop. Supposing typical values in the
flaring coronal loop: $n=10^{16} m^{-3}$, $l=5\times10^{7} m$, and
the cross-sectional radius is $r=5\times10^{6} m$ (Bray et al.
1991). Considering the period of preflare-VLPs is 1.9 - 47 min,
the electric current will be about $I=3.1\times10^{9} -
7.6\times10^{10}$ A.

Are these estimated electric currents reasonable? Many people
obtained the maximum current is in magnitude of $10^{12}$ A in an
active region during a solar flare derived from the vector
magnetograph observations (Canfield et al. 1993, Tan et al. 2006,
etc.). Reminding that an active region is always composed of
several decades or hundreds of plasma loops, it is reasonable to
suppose that the electric current will be in a magnitude of
$10^{10}$ A or lower in a single flaring plasma loop, especially
in the preflare phase.

Spangler developed a new method to observe the radio signal from a
remote quasar 3C228 when the emission passes by the solar limb
coronal loops, and obtained the electric current $10^{8}$ -
$10^{9}$ A in the coronal loops (Spangler 2007). In fact, during
the flare rising phase, QPPs are also existing frequently with
period from several seconds up to about 2 min (Simoes, Hudson, and
Fletcher 2015), which are shorter than the period of the above
preflare-VLPs. If we apply the same LRC oscillation mechanism to
explain them, a stronger electric current with magnitude of
$10^{10}$ - $10^{12}$ A can be obtained. It is natural because the
flaring plasma loop may become more unstable and generate stronger
electric currents during the flare rising phase.

The long-rising flares tend to be lack of preflare-VLP. It is
possible that the long-rising flare may have relatively slow
energy release and the flaring loops may disturb each other. These
mutual interferences make the dynamic processes become complex,
obscure the evidence of preflare-VLPs from the observations.

\section{Summary}

In brief, based on the analysis of the standard GOES SXR recorded
data, we discovered and confirmed that there were at least
one-third solar flares accompanying with preflare-VLPs, whose
typical periods were in the range of 8 - 30 min and durations in
the range of 1 - 2 hours. It is possible that this kind of
pulsations should be associated with some MHD oscillation modes,
similar to other QPPs occurring in the flare rising and postflare
phases. At the same time, the LRC-oscillation is also a favorable
mechanism to interpret the formation of preflare-VLP, which is
associated with current-carrying plasma loops. With
LRC-oscillation, the preflare-VLP may provide two important
information for understanding the solar explosions:

(1) The existence of preflare-VLPs indicates that electric
currents are generated in flaring plasma loops before the onset of
flares. The current-carrying loop may drive the plasma
instabilities, modulate SXR emission, and produce preflare-VLPs.
By studying the details of preflare-VLP, we may reveal the real
triggering mechanism of solar flares, the energy release, and
development of the source regions.

(2) The preflare-VLP can be regarded as a precursor of solar
flares. It is very simple to distinguish and extract a signal of
preflare-VLP from the GOES SXR observation data. And the duration
with 1 - 2 hours is long enough for us to response to the
influence of a powerful solar eruption. Therefore, preflare-VLP
can be regarded as a convenient precursory indicator to predict
the forthcoming of solar flares and the possibility of disastrous
space weather events and make an early warning.

Actually, besides the GOES SXR observation data, it is possible
that we can obtain much more information of the preflare-VLP from
the new generation, long-term continuous observations. For
example, the imaging observation of the Atmospheric Imaging
Assembly onboard NASA's satellite Solar Dynamic Observatory may
provide abundance imaging information of thermal emission in the
source regions with high spatial resolution (SDO/AIA, Lemon et al.
2012). Additionally, the Chinese Spectral Radioheliograph (CSRH,
Yan et al. 2009, now renamed Mingantu Spectral Radioheliograph,
MUSER) which can provide not only the nonthermal broadband
spectral structures, but also the variations of location and shape
in the source region with very high resolutions.

\acknowledgments The author thanks the referee for helpful and
valuable comments on this paper. we also thank the GOES teams for
providing the perfect observation data and the excellent software
for data analysis. This work is supported by NSFC Grant 11273030,
11661161015, 11221063, 11373039, 11573039, and 2014FY120300, CAS
XDB09000000.


\begin{thebibliography}{}
\bibitem[Alfven(1967)]{Alfven1967}Alfven, H. \& Carlqvist, P.: 1967, \textbf{SoPh}, 1, 220

\bibitem[Aschwanden(1987)]{Aschwanden1987}Aschwanden, M. J.: 1987, \textbf{SoPh}, 111, 113

\bibitem[Bloomfield(2012)]{Bloomfield2012}Bloomfield, D. S., Higgins, P. A., McAteer, R. T. J, \& Gallagher, P. T.: 2012, \textbf{ApJL}, 747, L41

\bibitem[Bray(1991)]{Bray1991}Bray, R. J., Cram, L. E., Durrant, C. J., \& Loughhead, R. E.: 1991, Plasma loops in the solar corona, New York: Cambridge University Press

\bibitem[Canfield(1993)]{Canfield1993}Canfield, R. C., de La Beaujardiere, J.-F., Fan, Y. H., et al.: 1993, \textbf{ApJ}, 411, 362

\bibitem[Chifor(2006)]{Chifor2006} Chifor, C., Mason, H. E., Tripathi, D., Isobe, H., \& Asai, A.: 2006, \textbf{A\&A}, 458, 965

\bibitem[Foullon(2005)]{Foullon2005}Foullon, C., Verwichte, E., Nakariakov, V. M., \& Fletcher, L.: 2005, \textbf{A\&A}, 440, L59

\bibitem[Garcia(1994)]{Garcia1994}Garcia, H. A.: 1994, \textbf{SoPh}, 154, 275

\bibitem[Georgoulis(2001)]{Georgoulis2001}Georgoulis, M. K., Vilmer, N., \& Crosby, N. B.: 2001, \textbf{A\&A}, 367, 326

\bibitem[Gonzalez(2014)]{Gonnlez2014}Gonzalez, H., I., Komm, R., Pevtsov, A., \& Leibacher, J. W.: 2014, \textbf{SoPh}, 289, 437

\bibitem[Harrison(1987)]{Harrison1987}Harrison, R. A.: 1987, \textbf{A\&A}, 182, 337

\bibitem[Huang(2010)]{Huang2010}Huang, X., Yu, D. R, Hu, Q. H., Wang, H. N., \& Cui, Y. M.: 2010, \textbf{SoPh}, 263, 175

\bibitem[Jiang(2015)]{Jiang2015}Jiang, M., Hu, D., Wang, X. G., et al.: 2015, \textbf{Nucl. Fusion}, 55, 083002

\bibitem[Kumar(2016)]{Kumar2016}Kumar, S., Nakariakov, V. M, \& Moon, Y.-J.: 2016, \textbf{ApJ}, 824, 8

\bibitem[Kupriyanova(2010)]{Kupriyanova2010}Kupriyanova, E.G., Melnikov, V.F.,  Nakariakov, V.M., \&  Shibasaki, K.: 2010, \textbf{SoPh}, 267, 329

\bibitem[Lemen(2012)]{Lemen2012}Lemen, J. R., Title, A. M., Akin, D.J., et al.: 2012, \textbf{SoPh}, 275, 17

\bibitem[Martin(1980)]{Martin1980}Martin, S.: 1980, \textbf{SoPh}, 68, 217

\bibitem[Nakariakov(2009)]{Nakariakov2009}Nakariakov, V. M., \& Melnikov, V. F.: 2009, \textbf{SSRv}, 149, 119

\bibitem[Nonweiler(1958)]{Nonweiler1958}Nonweiler, T.: 1958, \textbf{Natur}, 182, 468

\bibitem[Pick(2008)]{Pick2008}Pick, M., \& Vilmer, N.: 2008, \textbf{A\&ARv}, 16, 1

\bibitem[Roberts(1984)]{Roberts1984}Roberts, B., Edwin, P. M., \& Benz, A. O.: 1984, \textbf{ApJ}, 279, 857

\bibitem[Scafetta(2003)]{Scafetta2003}Scafetta, N., \& West, B. J.: 2003, \textbf{PRL}, 90, 8701

\bibitem[Shibata(2011)]{Shibata2011}Shibata, K., \& Magara, T.: 2011, \textbf{LRSP}, 8, 6

\bibitem[Simoes(2015)]{Simoes2015}Simoes, P. J. A., Hudson, H. S., \& Fletcher, L.: 2015, \textbf{SoPh}, 290, 3625

\bibitem[Spangler(2007)]{Spangler2007}Spangler, S. R.: 2007, \textbf{ApJ}, 670, 841

\bibitem[Svestka(1994)]{Svestka1994} Svestka, Z.: 1994, \textbf{SoPh}, 152, 505

\bibitem[Tan(2006)]{Tan2006}Tan, B. L., \& Huang, G. L.: 2006, \textbf{A\&A}, 453, 321

\bibitem[Tan(2006)]{Tan2010} Tan, B. L., Ji, H. S., Huang, G. L, Zhou, T. H., Song, Q. W., \& Huang, Y.: 2006, \textbf{SoPh}, 239, 137

\bibitem[Tan(2010)]{Tan2010} Tan, B. L., Zhang, Y., Tan, C. M., \& Liu, Y. Y.: 2010, \textbf{ApJ}, 723, 25

\bibitem[Tandberg-Hanssen(1988)]{Tandberg-Hanssen1988}Tandberg-Hanssen, E., \& Emslie, A.: 1988, The Physics of Solar Flares (Chapter 1),
Cambridge Uni Press

\bibitem[Tappin(1991)]{Tappin1991}Tappin, S. J.: 1991, \textbf{Astron. Astrophys. Suppl. Ser.}, 87, 277

\bibitem[Thomas(1985)]{Thomas1985}Thomas, R. J., Starr, R., \& Crannell, C. J.: 1985, \textbf{SoPh}, 95, 323

\bibitem[Tobias(2013)]{Tobias2013}Tobias, S. M., \& Cattaneo, F.: 2013, \textbf{Natur}, 497, 461

\bibitem[Wang(2011)]{Wang2011}Wang, T. J.: 2011, \textbf{Space Sci. Rev.}, 158, 397

\bibitem[Wesson(1997)]{Wesson1997}Wesson, J.: 1997, Tokamak(Chapter 7),
Clarendon Press, Oxford

\bibitem[White(2005)]{White2005}White, S. M., Thomas, R. J., \& Schwarts, R. A.: 2005, \textbf{SoPh}, 227, 231

\bibitem[Yan(2009)]{Yan2009} Yan, Y. H., Zhang, J., \& Wang, W., et al.: 2009, \textbf{EM\&P}, 104, 97

\bibitem[Yuan(2011)]{Yuan2011}Yuan, D., Nakariakov, V. M., Chorley, N., \& Foullon, C.: 2011, \textbf{A\&A}, 533, 116

\bibitem[Zaitsev(1998)]{Zaitsev1998}Zaitsev, V. V., Stepanov, A. V., Urpo, S., \& Pohjolainen, S.: 1998, \textbf{A\&A}, 337, 887

\bibitem[Zhang(2008)]{Zhang2008}Zhang, Y., Tan, B. L., \& Yan, Y. H.: 2008, \textbf{ApJ}, 682, L133

\bibitem[Zhang(2015)]{Zhang2015}Zhang, Y., Tan, B. L., \& Karlicky, M., et al.: 2015, \textbf{ApJ}, 799, 30

\end{thebibliography}
\end{document}